\documentclass{svproc}
\usepackage{bm,bbm,amsmath,amssymb,amsbsy ,subfigure,color,txfonts,multirow}
\usepackage{amsfonts}
\usepackage{array}
\usepackage[utf8]{inputenc}
\usepackage[framemethod=tikz]{mdframed}
\usepackage[colorlinks]{hyperref}
\hypersetup{
	bookmarksnumbered,
	pdfstartview={FitH},
	citecolor={blue}, 
	linkcolor={red},
	urlcolor={black},
	pdfpagemode={UseOutlines}
}

 \newcommand{\tr}[1]{\text{Tr}}
\newcommand{\ket}[1]{|#1\rangle}
\newcommand{\bra}[1]{\langle#1|}

\usepackage{url,graphicx,
multicol,footmisc}

\begin{document}
\mainmatter              % start of a contribution
\title{Information geometry of quantum resources}
\titlerunning{Information geometry}  % abbreviated title (for running head)
\author{Davide Girolami\inst{1}}
\authorrunning{Davide Girolami} % abbreviated author list (for running head)
 
%
\institute{Los Alamos National Laboratory,\\
 Theoretical Division, PO BOX 1663, Los Alamos 87545, NM, USA\\
\email{davegirolami@gmail.com},\\ WWW home page:
\texttt{https://sites.google.com/site/davegirolami/}
 }

\maketitle              % typeset the title of the contribution

\begin{abstract}
 I review recent works showing that information geometry is a useful framework to characterize quantum coherence and entanglement. Quantum systems exhibit peculiar properties which cannot be justified by classical
physics, e.g. quantum coherence and quantum correlations. Once confined to thought
experiments, they are nowadays created and manipulated by exerting an exquisite
experimental control of atoms, molecules and photons. It is important to identify and
quantify such quantum features, as they are deemed to be key resources to achieve
supraclassical performances in computation and communication protocols.
The information geometry viewpoint  elucidates the advantage provided by quantum superpositions in phase estimation. Also, it enables to link  measures of coherence and entanglement to observables, which can be evaluated in a laboratory by a limited number of  measurements.
\end{abstract}
\section{Introduction}
The possibility to prepare and manipulate even single, isolated atoms and photons makes possible to exploit quantum effects to speed-up information processing. In particular, the ability to create coherent superpositions of quantum states, enlightened by the iconic ``Schr\"{o}dinger's cat'' \cite{cat}, is the most fundamental difference between classical and quantum systems. Quantum information theory established coherence (the quantum label is omitted, from now on) as a key resource for obtaining an advantage in information processing \cite{nielsen,superrev,newmar,cohrev}.  Another critical property of quantum systems is entanglement, a kind of correlation which yields speed-up in information processing, as well as improves the precision of measurement devices \cite{entrev}. Yet,  to
quantify the coherence of quantum states, and the coherence consumed and created by quantum dynamics, access to the full state of a system is usually required. As its degrees of freedom are exponentially growing with the number of constituents, the task is computationally and experimentally challenging.\\
The works I here review employed ideas developed in classical and quantum information geometry \cite{amari,geo}, which visualize physical processes as paths on an abstract space, to develop  efficient strategies to evaluate  the coherence and the entanglement of a quantum state. The main result  is a certification
scheme, enabling to determine, by means of a limited number of measurements, the amount of coherence in a quantum state which is useful to phase estimation, the problem of reconstructing the value of an unknown parameter controlling the dynamics of a system. By generalizing the analysis to systems of many particles, an entanglement witness is obtained.    The proposal can be experimentally demonstrated by performing standard measurement procedures, being no a priori information available, thus outperforming methods involving state and
channel tomography, i.e. full reconstruction of the state of system. In fact, the test has been recently implemented in an all-optical setup via a network of Bell state projections \cite{speedexp}.  I also discuss an alternative experimental architecture which makes use of  spin polarization measurements, being suitable, for example, to Nuclear Magnetic Resonance (NMR) systems \cite{nmr}.

 \section{Making the usefulness of coherence manifest via information geometry}
 
 \subsection{Coherence as  complementarity between state and observable}
    The state of a finite dimensional quantum system is described by a self-adjoint semi-positive density matrix  $\rho, \rho=\rho^{\dagger}, \text{Tr}\{\rho\}=1, \rho\geq0$. Coherence can emerge whenever the system is prepared in a mixture of superpositions of two or more states. In other words, the density matrix representing the state is not diagonal with respect to a reference basis $\{h_i\}$, $\rho\neq \sum_i p_i \ket{h_i}\bra{h_i},  \langle h_i|h_j\rangle=\delta_{ij}, \sum_i p_i=1$. Surprisingly,  coherence  has been characterised as a resource  for information processing only recently. A consistent body of research identified the coherence of a quantum state  as  the ability  of a system to   break a symmetry generated by a Hamiltonian $H=\sum_i h_i \ket{h_i}\bra{h_i}$, i.e. a phase reference frame under a superselection
rule, where $H$ acts as the charge operator \cite{superrev,newmar}.  Coherence was then relabelled as {\it asymmetry}. Concurrent works proposed an alternative definition of coherence, as the distance of a state from the set of diagonal states in the reference basis \cite{cohrev}. In the following, I embrace the former interpretation.
An important question is  how to quantify asymmetry. A non-negative, contractive under noisy maps function of the state-observable commutator $[\rho, H]$ is arguably a good measure of asymmetry. Indeed,  whenever the state  is an eigenstate or a mixture of eigenstates of the observable, then it is diagonal in the Hamiltonian eigenbasis, which I assume here to be non-degenerate. However, it is desirable to link asymmetry to the performance in an information processing task. In other words, the asymmetry quantifier should be  the figure of merit of a procotol, benchmarking the usefulness of the system under scrutiny to complete the task. It is in fact possible to link asymmetry to the precision in phase estimation, as I explain in the next section.

\subsection{Quantum phase estimation}
 
 Metrology  is the discipline at the boundary between Physics and Statistics studying how to access information about a system by efficient measurement strategies and data analysis \cite{helstrom}.  Quantum metrology investigates how to improve the precision of measurements by employing quantum systems.  Results obtained in quantum metrology have found a use  in  interferometry, atomic spectroscopy, and gravitometry \cite{metrorev,toth}.   An important metrology primitive, as well as a frequent subroutine in computation protocols, is parameter estimation, which can be interpreted as a dynamical process. First, a probe system is prepared in an input state. Then, a controlled interaction imprints information about the parameter to estimate in the system state. Finally, a measurement is performed, to extract information about the parameter. A question to answer is what is the key property of the input state to maximise the precision of the estimation. It is known that quantum probes ourperform classical systems in a number of metrology tasks.  In particular, asymmetry is the key resource to phase estimation, a kind of parameter estimation where the perturbation of the system is described by a unitary dynamics. For the sake of clarity, I here review the protocol of parameter estimation, starting from the classical scenario \cite{helstrom}.
  A sample of independent measurement outcomes assigns values $x$ to a random variable $X$. The goal is to  construct a probability function $p_{\theta}(x)$.  
The exact value of the coordinate $\theta$ in the probability function space is not accessible. An  estimator $\hat\theta(x)$, and thus $p_{\hat\theta}(x)$, can be built yet.  The estimator is assumed to be unbiased.  This means that its  average value corresponds to the actual value of the parameter, $\int(\theta-\hat{\theta}(x))p_{\theta}(x)dx=0$.  The estimation precision is benchmarked  by the variance of the estimator $\hat{\theta}$. 
There exists a fundamental limit to the estimator performance.  One defines the optimal estimator $\hat{\theta}_{\text{best}}$ as the maximiser of the log-likelihood function $\max\limits_{\hat\theta}\ln l(\hat\theta|x)=\ln l(\hat{\theta}_{\text{best}}|x),  l(\hat\theta|x)\equiv p_{\hat\theta}(x)$.  The information about $\theta$ extracted by the measurements  is quantified by the score function $\frac{\partial \ln l(\theta|x)}{\partial \theta}$, which is  the rate of change of the likelihood function with the parameter value.   The second moment of the score is called the Fisher Information: 
\begin{equation}
F(\theta)=\int\left(\frac{\partial}{\partial\theta}\log p(x,\theta)\right)^2 p(x,\theta)dx.
\end{equation}
The Cram\'er-Rao bound establishes a lower limit to the variance of $\hat{\theta}$ in terms of such quantity,
\begin{equation}
V(p_{\theta},\hat{\theta})\geq\frac{1}{nF(\theta)},
\end{equation}
where $n$  is the number of repetitions of the experiment.  
Hence, the Fisher information is a figure of merit of the classical estimation protocol.\\
In the quantum scenario,  the state of the system   is represented by the density matrix $\rho_{\theta}$. Suppose to encode information about the parameter via a unitary transformation $\rho_{\theta}=U_{\theta} \rho_0 U^\dagger_{\theta}, U_{\theta}=e^{-i H \theta}$. The process corresponds to a path on the stratified manifold of the density matrices \cite{amari}. Assumed full knowledge of the Hamiltonian, but being the inital state unknown, what is the best strategy to extract the value of $\theta$?  One performs  a generalized positive operator value measure (POVM) $\{\Pi_x\}$ on the rotated  state $\rho_{\theta}$ \cite{nielsen}, where the $\{\Pi_x\}$  are the measurement operators  corresponding to the outcome $x$. One has $p_{\theta}(x)=\text{Tr}\{\rho_\theta\Pi_x\}$, and thus
\begin{eqnarray}\label{classf}
F(\rho_{\theta}):=\int dx \frac{1}{\text{Tr}\{\rho_{\theta} \Pi_x\}}\left(\text{Tr}\{\partial_\theta\rho_{\theta} \Pi_x\}\right)^2.
\end{eqnarray} 
 The optimal estimator, i.e. the most informative POVM, is a projection  into the eigenbasis of the symmetric-logarithmic derivative $L$,  which solves the equation $
\frac{\partial}{\partial\theta}\rho_\theta=\frac{1}{2}(\rho_\theta L+L\rho_\theta).$ Indeed, one has   $F(\rho_{\theta})\leq {\cal F}(\rho,H):= \text{Tr}\{\rho_{\theta}L^2\}$, where  ${\cal F}(\rho,H)$ is  the symmetric-logarithmic derivative quantum Fisher information (SLDF) \cite{helstrom}. Note that I omitted the parameter label for the state of the system, as the SLDF is independent of its value. The quantum version of the  Cram\'er-Rao bound  is then given by
 \begin{eqnarray}
 V(\rho,\hat{\theta}) \geq \frac{1}{n{\cal F}(\rho,H)}.
 \end{eqnarray}
 Given the spectral decomposition   $\rho=\sum_k\lambda_k\ket{k}\bra{k}$, the SLDF takes the expression
  \begin{equation}
\label{fisher}
{\cal F}(\rho,H)=\sum_{k<l}\frac{(\lambda_k-\lambda_l)^2}{2(\lambda_k+\lambda_l)}|\bra{k}H\ket{l}|^2,
\end{equation}
where each term in the sum is taken to be zero whenever $\lambda_i = \lambda_j$.\\
The quantity is well-known to the colleagues working in information geometry. It represents the norm related to the Bures metric, one of the quantum generalizations of the classical Fisher-Rao metric. These are special functions, being proven to be the unique Riemannian metrics which are contractive under quantum operations \cite{petzmono,geom}.  Consequently, the resource of the quantum protocol is the speed of evolution of the system undergoing the phase shift, i.e. how fast its state changes, as quantified by the SLDF.   I remind that for generic quantum operations  the most general expression of the quantum Fisher norms reads
\begin{eqnarray} 
||\partial_\theta\rho_\theta||_{f}^2&=&\sum\limits_{k,l}\frac{|\langle k(\theta)|\partial_\theta\rho_\theta|l(\theta)\rangle|^2}{\lambda_l(\theta)f(\lambda_k(\theta)/\lambda_l(\theta))}\nonumber\\
&=& \sum_k (d_\theta \lambda_k(\theta))^2 /4\lambda_k(\theta)  \nonumber\\
&+& \sum_{k< l} c_f(\lambda_k(\theta),\lambda_l(\theta))/2|\langle k(\theta)|\partial_\theta\rho_\theta |l(\theta)\rangle|^2,  
\end{eqnarray}
where $c_f(i,j)=(j f(i/j))^{-1}$, being $f$s the Chentsov-Morozova functions \cite{moro}.
The first term of the right hand side is the classical Fisher-Rao metric $\sum_k(d_{\theta}\lambda_k(\theta))^2/(4\lambda_k(\theta))$, which is the only one surviving for classical stochastic processes. On the other hand, for unitary transformations only the second, purely quantum term remains,  as only the eigenbasis of the state evolves. One has  $||\partial_\theta\rho_\theta||_{f}^2=f(0)/2 ||i[\rho_\theta,H] ||_{f}^2$. Then, the norm obtained by  fixing $f(x)={\cal F}(x)=(1+x)/2$ corresponds to the SLDF,  $||\partial_\theta\rho_\theta||_{F}^2={\cal F}(\rho, H)$.    In the more general case, when the quantum channel is not a unitary transformation, classical and quantum contributions co-exist.  \\

\begin{figure}[t]
 \centering
 \includegraphics[height=5cm,width=8cm]
 {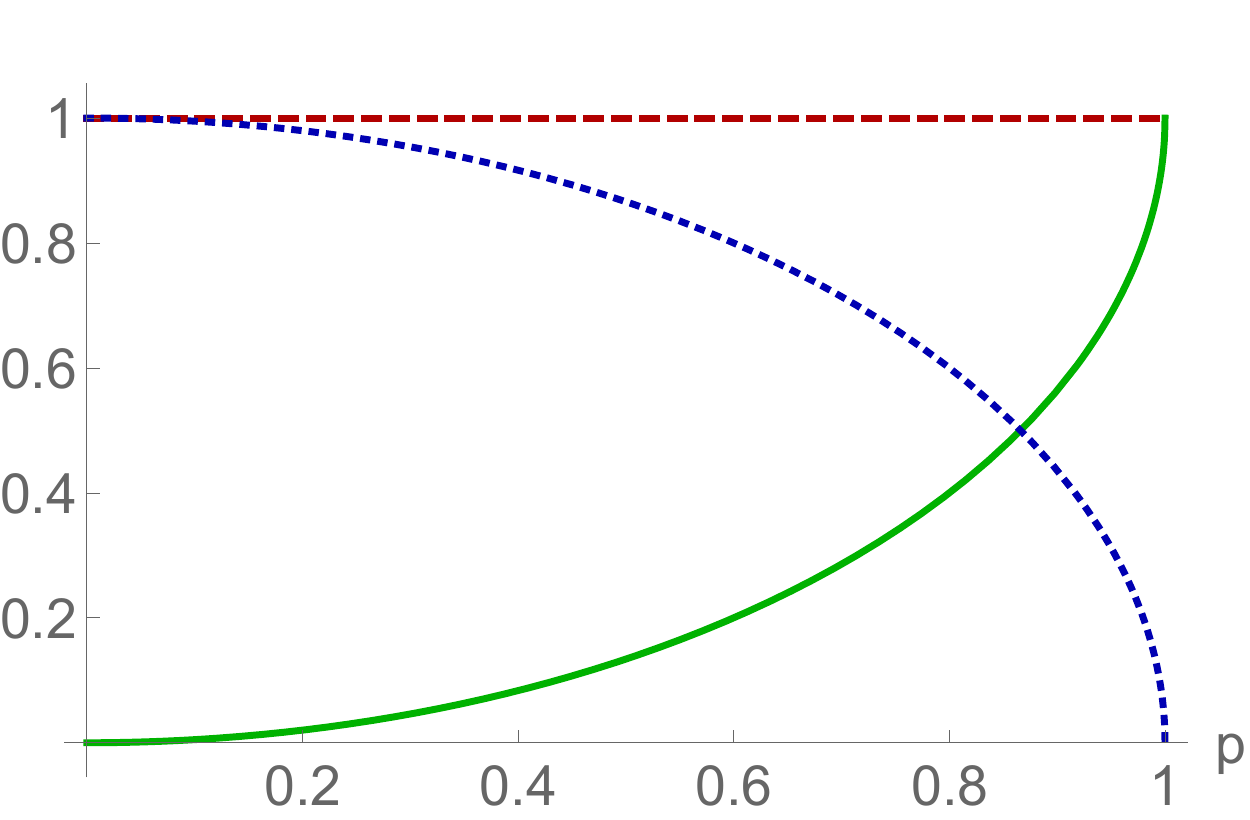}
\caption{Asymmetry and classical contribution to the variance are identified for the observable $\sigma_y=-i \ket{0}\bra{1}+i\ket{1}\bra{0}$
in the state $\rho =
(1 - p) I_2/2+ p \ket{\phi}\bra{\phi}, \ket{\phi}= 1/\sqrt 2(\ket{0}+\ket 1), p\in[0, 1]$. The red dashed
line is the variance,  while the green continuous curve is the SLDF. The blue
dotted curve  represents the difference between the two quantities, a classical mixedness measure. By varying the value of the noise parameter $p$, while the variance is constant, its quantum and classical components change. }\label{fig1}
\end{figure}

To summarize, the SLDF is a function of the commutator between state and Hamiltonian which has a natural interpretation as speed of the evolution of the system along a unitary dynamics. Also, it is a figure of merit of the phase estimation protocol. To verify that the SLDF is a consistent measure of asymmetry, therefore completing the characterization of asymmetry as an information processing resource, it has been proven that the SLDF satisfies a set of required properties \cite{ben}:
\begin{itemize}
\item[$\bullet$] The SLDF is upper bounded by the variance, ${\cal F}(\rho,H)\leq  V(\rho,H), V(\rho,H)=\text{Tr}\{\rho H^2\}-\text{Tr}\{\rho H\}^2$, where the equality is reached for  pure states. More precisely, the SLDF is the variance convex roof, ${\cal F}(\sum_i p_i \ket{\psi_i},H)=\inf\limits_{\{p_i,\ket{\psi_i}\}} \sum_i p_i V(\ket{\psi_i},H)$ \cite{toth2,yu}.
\item[$\bullet$] The SLDF is convex: ${\cal F}(p\rho_1+(1-p)\rho_2,H)\leq p{\cal F}(\rho_1,H)+(1-p){\cal F}(\rho_2,H)$.
\item[$\bullet$] For unitaries $U$, ${\cal F}(U\rho U^\dagger,H)={\cal F}(\rho,U^\dagger HU)$.
%\item[$\bullet$] The SLDF is non-increasing under quantum opeations which do not depend on the parameter: ${\cal F}(\Phi(\rho),H)\leq {\cal F}(\rho,H)$.
\item[$\bullet$] The SLDF is non-increasing under operations commuting with the phase shift, $ {\cal F}(\rho,H)\geq  {\cal F}(\Phi(\rho),H), \forall \Phi: [\Phi,U_\theta]=0$.  Note that  an even stronger constraint, contractivity on average under commuting operations, has been proven \cite{ben}.
\end{itemize}
Such properties are in fact met by all the regular quantum Fisher metrics, which are topologically equivalent to the SLDF \cite{gibilisco2}. Hence, they are legit measures of asymmetry.  While  the SLDF metric is the most employed one due to its operational interpretations in metrology, I observe that all the parent metrics may find, or have already found, their own operational interpretations. For example, the skew information $-1/2 \text{Tr}\{[\sqrt\rho, H]^2\}$ was introduced by Wigner and Yanase \cite{wigner}, and then generalized by Dyson \cite{wherl}, while discussing the implications of superselection rules in the measurement process.
\\
One may note that the variance enjoys both a simple expression and a close tie to experimental practice. However, it encodes a classical contribution   due to the mixedness of the state, such that it takes arbitrary high values even for states commuting with the observable, vanishing if and only if the state is an observable eigenstate. The variance is therefore not suitable to quantify asymmetry, apart from the pure state case. Conversely, the SLDF appears as the truly quantum contribution to the variance.  I clarify the interplay between variance  and SLDF by a simple example, discussed in Fig.~\ref{fig1}.

\section{Making coherence experimentally observable via information geometry}

\subsection{An observable lower bound to asymmetry} I here discuss how the information-geometric characterization of asymmetry as speed of evolution of a system yields  an experimentally friendly strategy to evaluate the asymmetry of an unknown state. 
I start by recalling a simple yet powerful algebraic result. Any degree $k$ polynomial function of a quantum state $f_k(\rho)$  equals the mean value of a self-adjoint operator $O_f$, measured on $k$ copies of the state: $f_k(\rho)=\text{Tr}(O_f  \rho^{\otimes_k})$ \cite{paz,brun}. This is useful because in Quantum Mechanics observables, i.e. measurable quantities, are represented by self-adjoint operators.  Then, searching for polynomial approximations is a convenient strategy to overcome expensive state tomography when one wants to determine non-directly observable quantities, e.g. quantifiers of non-linear properties as coherence. \\
Measuring the corresponding observable $O_f$ is not guaranteed  to be practicable. However, this is provably possible for the simplest case of quadratic polynomials, e.g.  the overlap between two states $\text{Tr}\{\rho \sigma\}$.  By selecting the swap operator $V(\phi_1\otimes\phi_2)=\phi_2\otimes\phi_1, \forall \phi_{1,2},$ as probe observable acting on the tensor product of two copies of the system Hilbert space, one has $\text{Tr}\{\rho \sigma\}=\text{Tr}\{V (\rho\otimes\sigma)\}, \forall \rho,\sigma$. The swap can be measured by single qubit interferometry. Two copies of the system of interest, or two different degrees of freedom of a single replica, say spin and linear momentum, are prepared in the states $\rho, \sigma$. They are correlated by a controlled-swap gate to an ancillary qubit in the initial state $\alpha_{\text{in}}$, which acts as the control qubit. The mean value of the swap is then encoded in the polarisation of the output state of the ancilla, $\text{Tr}\{\alpha_{\text{out}}\sigma_z\}=\text{Tr}(\alpha_{\text{in}}\sigma_z)\text{Tr}\{V (\rho\otimes\sigma)\}$, where $\sigma_{x,y,z}$ are the spin-1/2 Pauli operators. (See the scheme in \cite{me}.) A shortcoming of the scheme is that it is currently hard  to engineer high fidelity controlled-swaps. The minimal three qubit architecture has been experimentally demonstrated only recently \cite{fredkin}. It is nevertheless possible to overcome the problem whenever the system of interest displays a partition in $N$ subsystems $\{A_i\}, i=1,\ldots,N$, e.g. it is an $N$-qubit computational register \cite{alves}.  By observing that the swap is factorizable, $V_{A_1\ldots A_N}=\otimes_i V_{A_i},$ one has $\text{Tr}\{\rho_{A_1\ldots A_N} \sigma_{A_1\ldots A_N}\}=\text{Tr}\{\otimes_i V_{A_i} (\rho_{A_1\ldots A_N}\otimes \sigma_{A_1\ldots A_N})\}$. The state overlap is then obtained by a collective detection of $O(2N)$ local observables on  two copies of the $N$-partite register. The scheme performs exponentially better than the $O(4^{N})$ measurements needed to state tomography.  Note that state reconstruction also needs an equivalent number of system copies to perform the measurements.  This is  an experimentally exhausting task
already for few qubits, even without considering the exposure to error sources
affecting the detection, which arguably grows with the number of measurements. Moreover, one usually  finds that protocols involving a limited number of measurements are faster and easier to control.
Indeed, the very same existence of full-fledged research lines is devoted to avoid
state tomography, e.g. the works in compressed sensing and state discrimination. 
  In the qubit case, local Bell singlet projections are sufficient to evaluate the swap, $V=1-2 P_-, P_-=\ket{\phi^-}\bra{\phi^-},\ket{\phi^-}=1/\sqrt2(\ket{01}-\ket{10})$.  They are implemented by beam splitter interactions between each subsystem $A_i$ copy pair, and single-site polarization detections.  A further alternative scheme relying on correlating the system with an array of ancillary qubits has been proposed \cite{me}. 

I now show that picking the SLDF as a quantifier of asymmetry is useful for experimental practice. No measure of asymmetry can take the form of self-adjoint operators, as it is a non-linear property of a state. Yet, it is possible to construct a geometric lower bound to the SLDF, and in general to any regular quantum Fisher metric (up to a factor), which is a function of observables \cite{speedexp,me,ent}. By employing the Hilbert-Schmidt norm $||A||_2=\sqrt{\text{Tr}\{AA^{\dagger}\}}$, one has
\begin{eqnarray}
  {\cal S}_\theta(\rho,H)&\leq& {\cal F}(\rho,H),  \\
{\cal S}_\theta(\rho,H)&=&||U_\theta \rho U_\theta^{\dagger} -\rho ||_2^2/ (2\theta^2)=(\text{Tr}\{\rho^2\}-\text{Tr}\{\rho U_\theta \rho U_\theta^{\dagger}\})/\theta^2.\nonumber
\end{eqnarray} 
 
The proof of the result is given in Ref.~\cite{speedexp}.  Thus, a lower bound to asymmetry is given as a function of purity and overlap, as well as the parameter $\theta$, whose value is experimentally controllable. As discussed before, quadratic polynomials are directly measurable, provided two system replicas $\rho_{1,2}\equiv \rho$. One has $\text{Tr}\{\rho^2\}=\text{Tr}\Big\{V(\rho_1\otimes\rho_2)\Big\}, \text{Tr}\Big\{\rho U_\theta \rho  {U_\theta^{\dagger}}\Big\}=\text{Tr}\Big\{V\Big(\rho_1\otimes U_{\theta}\rho_2{U_{\theta}^{\dagger}}\Big)\Big\}.$
 The result is valid for arbitrary input states. In particular, the method is suitable for large scale detection of asymmetry, as it requires a limited number of measurements regardless the dimension of the system under scrutiny.

\subsection{Asymmetry witnesses Entanglement}
The notion of asymmetry (coherence) can be applied to systems which display a structure, e.g. described a partition ${\cal S}\rightarrow \{{\cal S}_i\}$. The partition is usually determined by the particulars of the many-body system, as the spatial separation between the parts.  One may ask what implies that the state of a multipartite system has asymmetry, and what a measure of asymmetry can reveal about the interdependence between the system parts. In spite of being a basis-dependent feature, asymmetry is affected by quantum correlations, which are basis independent properties of multipartite systems. Indeed, by measuring the observable ${\cal S}_\theta(\rho,H)$, entanglement between the particles can be witnessed. There are several entanglement indicators written in terms of the Fisher information, among the many strategies proposed to detect entanglement \cite{toth3}. In particular, one has that, for $N$ qubits, if $\bar{{\cal F}}(\rho)=1/3({\cal F}(\rho, J_x)+{\cal F}(\rho, J_y)+{\cal F}(\rho, J_z))>N/6 , J_{x(y,z)}=\sum_i 1/2\sigma_{x(y,z)}^i,  \sigma_{x(y,z)}^1=\sigma_{x(y,z)\ 1}\otimes I_{23},\sigma_{x(y,z)}^2=I_{1}\otimes \sigma_{x(y,z)\ 2}\otimes I_{3},\sigma_{x(y,z)}^3=I_{12}\otimes \sigma_{x(y,z)\ 3}$, %$\chi_H=N/{\cal F}_H(\rho)<1$ or 
then the state is entangled \cite{smerzi}.  Also, a so called $k$-separable state of $N$ qubits cannot satisfy ${\cal F}(\rho, J_{x(y,z)}) \geq (nk^2 + (N-nk)^2)/4 $, 
where $n = \lfloor \frac{N}{k} \rfloor$. For example, given $N=3$, one has $k=1 \Rightarrow {\cal F} \geq 3/4$ and $k=2 \Rightarrow {\cal F} \geq 5/4$. These bounds represent a witness of entanglement and genuine tripartite entanglement, respectively. Then, the asymmetry lower bound, for additive spin Hamiltonians, is also an entanglement witness:
\begin{eqnarray}\label{witness}
\bar{{\cal S}}_{\theta}(\rho)&=&1/3({\cal S}_\theta(\rho, J_x)+{\cal S}_\theta(\rho, J_y)+{\cal S}_\theta(\rho, J_z))> N/6\nonumber\\
{\cal S}_\theta(\rho, J_{x(y,z)}) &\geq& 1/4(nk^2 + (N-nk)^2).
\end{eqnarray}
It is remarkable that superlinear scaling of asymmetry in multipartite systems  witnesses  entanglement in action,  not just non-separability of the state. In other words, it detects when entanglement provides a tangible advantage, making the state evolve faster under a phase encoding evolution.

\section{Experimental implementation}

\subsection{Detecting asymmetry via Bell state projections}
 
The result calls for experimental demonstration in standard quantum information testbeds.  Let us apply the scheme to simulate the detection of coherence and entanglement in a two-qubit state. A similar experiment has been recently implemented in optical set-up \cite{speedexp}. For convenience, I choose a Bell-diagonal state  $\rho^p=p(\ket{\psi}\bra{\psi})+(1-p)/4 I_4, \ket{\psi}=1/\sqrt2 (\ket{00}+\ket{11}), p\in[0,1],$ as a probe state for the test. This allows for investigating the behavior of the asymmetry lower bound and entanglement witness in the presence of noise. Two copies of a Bell diagonal state $\rho^p_{A_1B_1},\rho^p_{A_2B_2}$ are prepared. One can verify the behaviour of asymmetry with respect to a  set of spin observables $J_{K}=\sum_{i=A,B} j_K^i, j_K^A=j^K_A\otimes I_{B},j_K^B=I_{A}\otimes j^K_B, j_K=1/2 \sigma_{K=x,y,z},$ by  varying the value of the mixing parameter $p$.   This is done by implementing the unitary gate $U^A_{K,\theta}\otimes U^B_{K,\theta}, U_{K,\theta}=e^{-i J_K \theta}, $ on a copy of the state.   One then needs to evaluate the purity of the state of interest and an overlap with a shifted copy  after a rotation has been applied.  Note that to evaluate the purity, no gate has to be engineered.   For optical setups, one rewrites the two quantities in terms of projections on the antisymmetric subspace. The swap acting on the register $A_1B_1A_2B_2$ is the product of two-qubit swaps on each subsystem, $V_{A_1B_1A_2B_2}=V_{A_1A_2}\otimes V_{B_1B_2}$. Also, for two qubit swaps, one has $V_{12}=I-2 P^-_{12} , P^-_{12}=\ket{\psi}\bra{\psi}_{12}, \ket{\psi}=1/\sqrt2(\ket{01}-\ket{10})$.  Thus, the observables $O_i$ to be measured are
\begin{eqnarray}\label{meas}
O_1&=&P^-_{A_1A_2}\\
O_2&=&P^-_{B_1B_2}\nonumber\\
O_3&=&P^-_{A_1A_2}\otimes P^-_{B_1B_2} \nonumber\\
\text{Tr}\{\rho^2\}&=&1+ 4\text{Tr}\{O_3 \rho_{A_1B_1}\otimes \rho_{A_2B_2}\}-2\text{Tr}\{O_1 \rho_{A_1B_1}\otimes \rho_{A_2B_2}\}\nonumber\\
&-&2\text{Tr}\{O_2 \rho_{A_1B_1}\otimes \rho_{A_2B_2}\}\nonumber\\
\text{Tr}\{\rho_{A_1B_1} U_\theta \rho_{A_2B_2} U_\theta^{\dagger}\}&=&1+ 4\text{Tr}\{O_3 \rho_{A_1B_1}\otimes U_\theta \rho_{A_2B_2} U_\theta^{\dagger}\}-2\text{Tr}\{O_1 \rho_{A_1B_1}\otimes U_\theta \rho_{A_2B_2} U_\theta^{\dagger}\}\nonumber\\
&-&2\text{Tr}\{O_2 \rho_{A_1B_1}\otimes U_\theta \rho_{A_2B_2} U_\theta^{\dagger}\}.\nonumber
\end{eqnarray}  
  No further action is necessary to verify the presence of entanglement through the witnesses in Eq.~(\ref{witness}). For $N=2$, one has to verify  ${\cal S}_\theta(\rho, J_{K}) \geq 1/2$ and ${\cal\bar{S}}_\theta(\rho)>1/3$.  The results are summarised  in Table~\ref{theory}.

   \begin{table}
\centering
\begin{tabular}{|c|c|c|c|}
\hline
 $J_k $&$J_x$& $J_y$  &$ J_{z}$\\
\hline
 ${\cal F}(\rho^p_{AB}, J_K)$&  $2 p^2/(p+1)$ &    $0$
&    $2p^2/(p+1)$ \\
\hline
${\cal S}_\theta(\rho^p_{AB}, J_K)$&   $p^2 \sin\theta^2/ \theta^2 $& $0$
&  $  p^2 \sin\theta^2/\theta^2  $   \\
\hline
${\cal F}_(\rho^p_{AB}, J_K)>1/2$ &$p > 0.640388$& $/$
 &$p > 0.640388$   \\
 \hline
  $ {\cal S}_{\pi/6} (\rho^p_{AB}, J_K)>1/2$& $p > 0.74048$&$/$& $p > 0.74048$      \\
 \hline
\multicolumn{1}{|c|}{}&\multicolumn{3}{c|}{}\\
  ${\cal\bar{F}}(\rho^p_{AB})>1/3,\bar{{\cal S}}_{\pi/6}(\rho^p_{AB})>1/3$ & \multicolumn{3}{c|}{
     $p > 0.\bar{3}, p > 0.427517$}  \\[1ex]
\hline
\end{tabular}
\vspace{15pt}
\caption{Theoretical values of the SLDF, the lower bound,  and the conditions witnessing entanglement, for the spin observables $J_{K}$, in $\rho^p_{AB}$,  by fixing  $\theta=\pi/6$. The lower bound is an entanglement witness almost as efficient as the quantum Fisher information, being not able to detect metrologically useful  entanglement for $p\in[0.\bar{3}, 0.427517 ]$. Note that the state is entangled for $p>1/3$.}
\label{theory}
\end{table}
  %\begin{figure*} 
 %\centering
%\subfigure[]{\includegraphics[width=.3\textwidth]{x.pdf}}
%\subfigure[]{
%\includegraphics[width=.3\textwidth]{y.pdf}}
%\subfigure[]{\includegraphics[width=.3\textwidth]{z.pdf}}
%\caption{(Colors Online) -- Evaluation of quantum coherence in the state $\rho=\rho^p_{AB}$ with respect to the observables $K=J_x, J_y,J_z$ (figures (a), (b) and (c) respectively) as a function of the mixing parameter $p$. The blue dotted line is the quantum Fisher information, here showed for reference, the red dashed line %is the bound ${\cal }O_{K}(\rho)$, the red continuous %line is the approximated quantity ${\cal %}O^{\text{ap}}_{K}(\rho)$ obtained by imposing %$\theta=\pi/6$, and the yellow shadow represents the %error bar, bounded by the continuous yellow lines %representing the extreme values of ${\cal %}O^{\text{ap}}_{K}(\rho)\pm \Delta{\cal %}O^{\text{ap}}_{K}(\rho) $. }
%\label{plot}
%\end{figure*} 

\subsection{Alternative scheme}

Let us suppose that the expectation values of spin magnetizations are measurable, e.g. as it happens in an  NMR system \cite{nmr}. Given a $n$-qubit register, without the possibility to  perform projections, a full state reconstruction would require $2^{2n}-1$ measurements.
However,  one can always retrieve the value of the overlap of any pair of states $\text{Tr}\{\rho\sigma\}$ by the same amount of measurements required by the tomography of a {\it single} state.  In our case, by retaining the possibility to implement two state copies  (our bound is a polynomial of degree two of the density matrix coefficients), one can evaluate the purity and the overlaps with  $2^{2n}-1$ measurements, avoiding to perform tomography on both the states.  There is also a further advantage: there is no need to apply any additional (controlled or not) gate to the network.  Let us suppose one wants to extract information about the asymmetry and the entanglement of a three-qubit state $\rho_{ABC}$. I assume for the sake of simplicity that our state has an X-like density matrix, i.e. it is completely determined by 15 parameters. Thus, 15 measurements are sufficient for state reconstruction. The parameters are the expectation values of magnetization measurements: $\rho^X_{ABC}=1/8(I_8+\sum_i\text{Tr}\{\rho^X_{ABC} m_i\})$, where:
\begin{eqnarray}\label{meas}
m_1&=& 4 \sigma_A^z\otimes I_B\otimes I_C\\
m_2&=&4 I_A\otimes\sigma^z_B\otimes I_C\nonumber\\
m_3&=&4I_A\otimes I_B\otimes\sigma^z_C\nonumber\\
m_4&=&16 \sigma_A^z\otimes\sigma^z_B\otimes I_C\nonumber\\
m_5&=&16 I_A\otimes\sigma^z_B\otimes\sigma^z_C\nonumber\\
m_6&=&16\sigma_A^z\otimes I_B\otimes\sigma^z_C\nonumber\\
m_7&=& 64\sigma_A^z\otimes\sigma_B^z\otimes \sigma_C^z\nonumber\\
m_8&=&64\sigma_A^x\otimes \sigma_B^x\otimes \sigma_C^x\nonumber\\
m_9&=& 64\sigma_A^x\otimes \sigma_B^x\otimes \sigma_C^y\nonumber\\
m_{10}&=&64\sigma_A^x\otimes \sigma_B^y\otimes \sigma_C^x\nonumber\\
m_{11}&=&64\sigma_A^y\otimes \sigma_B^x\otimes \sigma_C^x\nonumber\\
m_{12}&=&64\sigma_A^y\otimes \sigma_B^y\otimes \sigma_C^x\nonumber\\
m_{13}&=& 64\sigma_A^y\otimes \sigma_B^x\otimes \sigma_C^y\nonumber\\
m_{14}&=& 64\sigma_A^x\otimes \sigma_B^y\otimes \sigma_C^y\nonumber\\
m_{15}&=& 64\sigma_A^y\otimes \sigma_B^y\otimes \sigma_C^y.\nonumber
\end{eqnarray}

As the swap  is factorizable,  any overlap $\text{Tr}\{\rho^X_{ABC} \sigma_{ABC}\}$ is fully determined by the very same measurements $\{M_i=m_i\otimes m_i\}$, regardless of the density matrix of $\sigma$. 
  Also, the measurements to perform are independent of the specific observable $J_K$. By noting that $e^{i \sigma_i\theta} \sigma_j e^{-i \sigma_i \theta}= \cos{(2 \theta)}\  \sigma_j-\sin{(2\theta)}\  \sigma_k \epsilon_{ijk},$
one finds that for any overlap  
\begin{eqnarray}\label{measurements}
\text{Tr}\{\rho_{ABC}^p U_\theta\rho^X_{ABC} U^{\dagger}_\theta\}&=&1/8\text{Tr}\{\rho^X_{ABC}\otimes U_\theta\rho^X_{ABC} U^{\dagger}_\theta M_i\}.
\end{eqnarray}
 The argument can be generalized to states of any shape and  dimension.

\section{Conclusion}
I here presented an overview of recent works employing information geometry concepts to characterize the most fundamental quantum resources, i.e. coherence and entanglement.   It is proven that the SLDF is an asymmetry measure, quantifying the coherence of a state with respect to the eigenbasis of a Hamiltonian $H$. The results holds for  any regular quantum Fisher information metric.
Furthermore, the SLDF, as well as the parent metrics, is lower bounded by a function of observable mean values. Such geometric quantity can be then related to experimentally testable effects, i.e. statistical relations in measurement outcomes. 
     When the Hamiltonian is an additive spin operator generating a many-body system dynamics, the lower bound is an entanglement witness. Experimental schemes  to detect asymmetry and entanglement, which are implementable with current technology, have been described.\\
   It would be interesting to investigate whether the sensitivity of a system to  non-unitary but still quantum evolutions is also linked to the presence of quantum resources. The scenario is certainly closer to realistic experimental practice, where the system is disturbed by uncontrollable error sources, accounting for imperfections in state preparation and dynamics implementation. I also anticipate that employing the information geometry toolbox may critically advance our understanding of genuinely quantum information processing. For example, by establishing fundamental geometric limits and optimal strategies to the control of quantum systems. 

\paragraph{Notes and Comments.}
I thank the organizers and the participants of the IGAIA IV conference for the hospitality and the stimulating discussions. This work was supported by the Los Alamos National Laboratory, project 20170675PRD2. Part of this work was carried out at the University of Oxford, supported by the UK Engineering and Physical Sciences Research Council
(EPSRC) under the Grant No. EP/L01405X/1, 
and by the Wolfson College.

\end{document}